# Taxi-based Emergency Medical System

Li-Yi Lin, llin34@jhu.edu

**Abstract:** In case of a severe accident, the key to saving lives is the time between the incident and when the victim receives treatment from the first-responders. In areas with well-designed emergency medical systems, the time for an ambulance to arrive at the accident location is often not too long. However, in many low and middle-income countries, it usually takes much longer for an ambulance to arrive at the accident location due to lack of proper services. On the other hand, with ubiquitous wireless connectivity, and emergence of radio-based taxis, it seems feasible to build a low-cost emergency response system based on taxi service. In this report, we explore one such solution for deployment of a taxi-based emergency response systems using reinforcement learning.

## 1. Introduction

When a severe accident happens, the key of saving people's life is the time for the victim to receive a prompt treatment. The shorter the time, the higher chance that the wounded can be saved. The time is also called golden hour[1]. In areas that have well-designed emergency medical system (EMS) with enough ambulances, the time for an ambulance to arrive the accident location would not be too long. However, in other areas such as low- and middle-income countries (LMICs), it will take a longer time for the ambulance to arrive the accident's location due to the lack of ambulances (Bhalla, 2015).

Nowadays, technologies have already created a new way for taxi services. The dispatch system can monitor the location of each taxi and call the nearest taxi to passengers according to the estimated traffic time or distance. With more and more ubiquitous wireless Internet accessibility and affordable smartphones, this kind of taxi service is also applicable in the LMICs. If taxi drivers are trained to give a prompt treatment to the victims, then the taxi dispatch system will be a good complementary to the emergency medical system and save more life.

However, the traffic situation might be changed because of a car accident (or other events that will change the traffic situation) and using the current traffic information to predict the arrival time is not the best way. We need to pick the taxi that can arrive the accident's location with the minimum time among all the available taxis while the traffic situation is changing. But how can we make a right decision? Nowadays, machine learning techniques have successfully applied to many research areas and real world applications, such as a self-driving car. For the problem we mentioned above, supervised learning techniques might not be appropriate due to the lack of labeled training data. One technique that can continuously learn the policy from experience is called reinforcement learning.

Reinforcement learning algorithm can gradually learn a policy to a problem from each time it makes a decision. When a decision is made, a corresponding feedback (reward) will be learned and the algorithm can adjust its policy according to it (Sutton & Barto, 1998). Reinforcement learning has also successfully been applied to many areas, such as robotics and Game playing. For example, the Google AlphaGo beat a Go world champion, Lee Sedol, in March 2016[2], which is a big breakthrough in artificial intelligence. AlphaGo adopted reinforcement learning algorithm as part of its game playing intelligence and learns the strategy from each game. It means that the reinforcement learning is a really powerful learning algorithm for solving complicated problems. Thus, this method is more suitable for the original problem.

---

[1] https://en.wikipedia.org/wiki/Golden_hour_(medicine)
[2] https://en.wikipedia.org/wiki/AlphaGo_versus_Lee_Sedol

For the structure of this project report, we will first describe a traffic simulator built by ourselves that incorporate the intelligent driver model (Intelligent driver model, n.d.) and use real world map data to create a simulated environment. Then we perform experiments to see if a navigation based on average traffic time can help the called taxi to arrive the crash location faster. We will also discuss the experiment result and the possible application of reinforcement learning algorithms for this problem in the last section.

## 2. Traffic Simulator

Since we don't possess real-world traffic data for training the reinforcement learning algorithm, we plan to use a traffic simulation program to generate simulated traffic. There is one traffic simulation program called "TRansportation ANalysis and SIMulation System (TRANSIMS)" (TRANSIMS, n.d.) (Transims Open Source, n.d.). This program was created by a project originally funded by U.S. governments[3] and it later became an open source project[4]. Although this program provides an extremely detailed simulated environment, it will take 6 months to a year for people to become an effective TRANSIMS user according to the tutorial data given by Argonne National Laboratory[5]. In addition to using the program, it also needs to extra modules that transmit the information of the simulated environment to the learning algorithm and an API for the algorithm to call for taxis to a crash location. Inspired by a light traffic simulator (Volkhin, n.d.) and under the limited time and human resource, we then built a light version of the traffic simulation program.

The goal of this traffic simulation program is to provide an environment that mimics the real world situation. This program can be divided into two parts: one is the map construction and the other is the traffic simulation using the "Intelligent driver model." For this report, we used the map of Washington D.C. as an example.

### 2.1. Map Construction
First, we will describe how the map is built. The simulation program parses the shape file that stores the GPS coordinates of roads and intersections in Washington D.C.[6] and reconstructs the map.

### 2.1.1. Intersection
The shape file directly provides information about the intersections in the area. At each intersection, there will be a traffic light controller determining when cars can go across the intersection from one road to the other that are connecting to the intersection. Every traffic light controller changes its signals every certain length of time. The length of time for every controller to change its signals ranges from 15 to 30 seconds.

---

[3] The Federal Highway Administration, the Federal Transit Administration and the Office of the Assistant Secretary for Transportation Policy of the United States Department of Transportation, and the Environmental Protection Agency.
[4] https://sourceforge.net/p/transims/code/HEAD/tree/
[5] TRANSIMS training courses, https://www.tracc.anl.gov/index.php/transims-training-course
[6] Shape file of Washington D.C. at District of Columbia Open Data website: http://opendata.dc.gov/datasets/e8299c86b4014f109fedd7e95ae20d52_61?geometry=-77.439%2C38.826%2C-76.573%2C38.986&mapSize=map-maximize

## 2.1.2. Road

When a road record is read, the program then connects the road's end points with intersections that it can reach. If a road connects two intersections, then the program will keep it on the map. For every road record, we create two roads with opposite directions. Every road consists of two lanes so that cars and taxis can switch to different lanes when they found a quicker one. See Figure 1 for the road construction. We also put a speed limit on all roads and it is 60 km per hour in our setting. Every car and taxi cannot drive faster than the speed limit of the road where it is driving on.

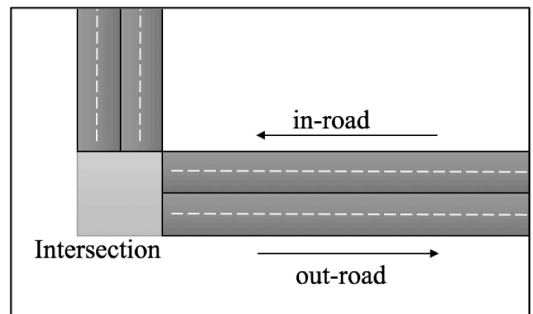

**Figure 1.** Every road has two lanes. In-road and out-road are always paired together. Every road has two lanes.

In the real world, there will be some roads that have more traffic. We called those roads "major road." For simplicity, we think major roads are usually longer. So, if a road is longer than a certain length (200 meters in our setting), the simulation program will mark it a major road. In addition, if a road connects two major roads, it will also be marked as a major road.

We adopted Poisson point process to generate the traffic. Major roads have a higher $\lambda$ value that is the intensity or the expected number of Poisson points in some time period so that there will be more traffic on them. The constructed map is shown in Figure 2.

## 2.1.3. Source and Sink Point

Cars or taxis usually drive from their current location to their destinations. To simulate this process, we introduced source and sink points. A source point is a location where a car can be generated and put on the map. This process is like that a car drives out from a parking lot or a street parking grid. A sink point is a destination for cars and taxis. When a car arrives its destination (sink point), it will be removed from the map. When a taxi arrives its destination, a new destination will be assigned to it. Every car and taxi will be assigned a random sink point. During the simulation, cars and taxis will try to arrive their destinations.

We now describe how we choose locations for source and sink points. Every intersection has in-roads and out-roads (see Figure 1). The "in-road" means that cars will go toward the intersection. The "out-road" means that cars will drive away from the intersection. In other words, the in-road's direction is pointing to the intersection. For an intersection that has only one in-road and out-road, we put a source and sink point on it since it is an end point and we see it as an edge of the map.

The program will also assign source and sink points on roads. The program will start at a random intersection and perform bread first search on the entire map. When it found a closed area that is surrounded by roads, source and sink points will be assigned to a random position for the last closed road except the first and last 10% of the road to prevent it from being too close to the intersection. See Figure 3 for an example.

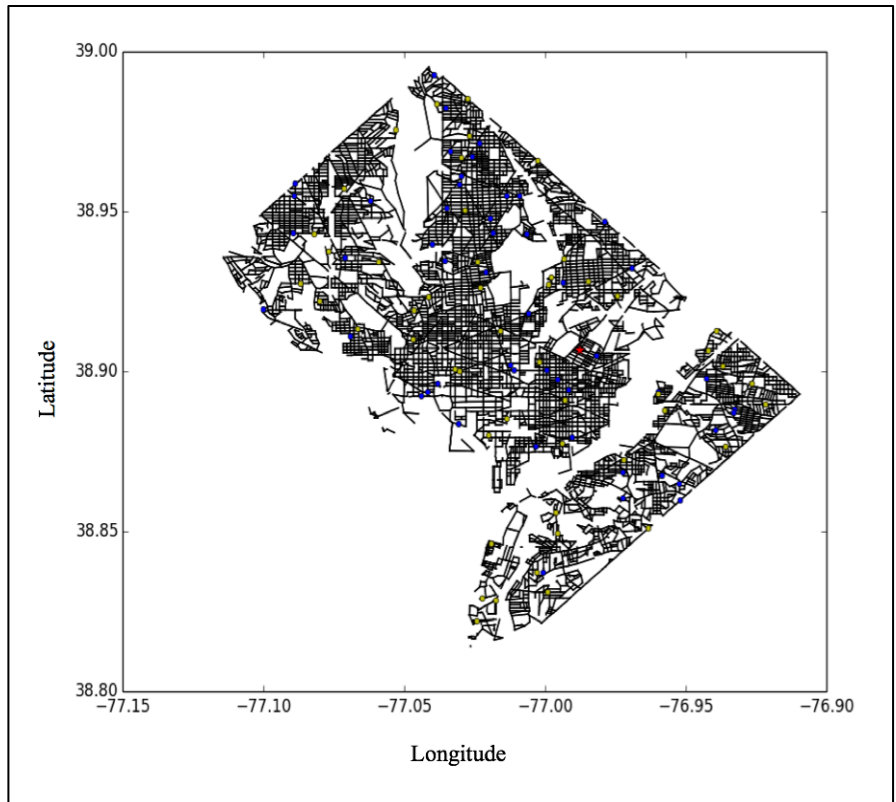

**Figure 2.** The constructed map using the shape file of Washington D.C. The numbers on the edge are the latitude and longitude. The blue and yellow points are cars and taxis respectively. The red start is the location for a crash event.

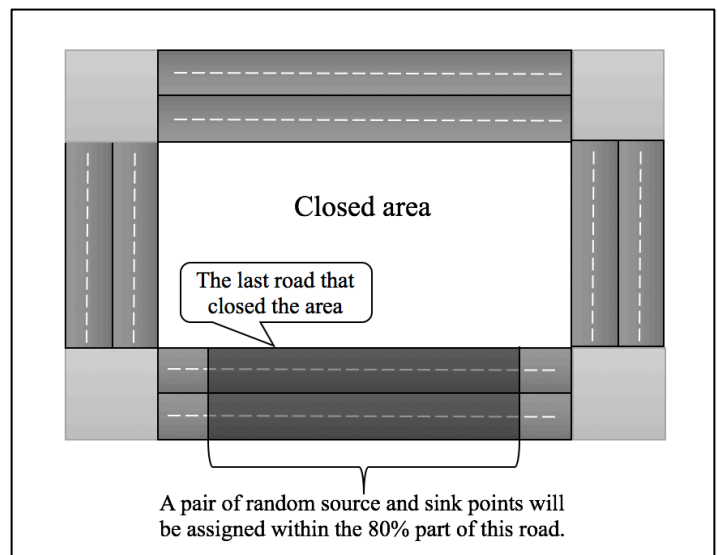

**Figure 3.** The demonstration of where source and sink points can be assigned to the road that closed an area.

### 2.1.4. Map Visualization

In order to make the map look like a real one, we also added a satellite map image downloaded from Google Map API to the background. Although the roads do not fully match the satellite map because the program uses the averaged center position for the intersections, the shape of constructed map is still very similar to the real one. See Figure 4 for the map visualization.

In Figure 4, the red and blue rectangles indicated the cars running on the left and right lanes respectively. The yellow rectangles represent the taxis. The pink rectangle is the taxi that has the shortest estimated time to the crash location when the crash happens. The yellow star is the crash event location. The green lines are the major roads. The size of the cars and taxis are much bigger than the real size relative to that of the map. We enlarge the size for people to see the cars and taxis more easily.

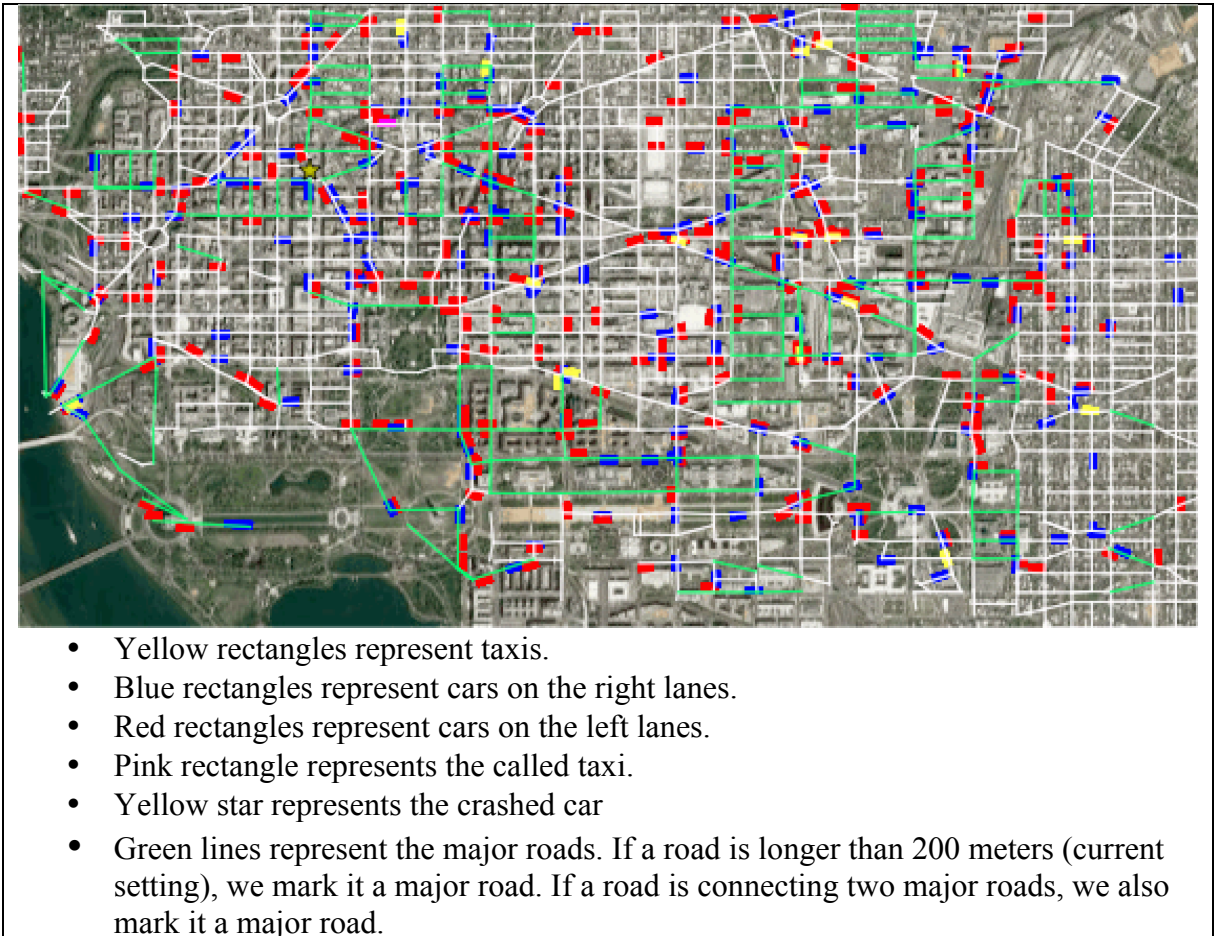

- Yellow rectangles represent taxis.
- Blue rectangles represent cars on the right lanes.
- Red rectangles represent cars on the left lanes.
- Pink rectangle represents the called taxi.
- Yellow star represents the crashed car
- Green lines represent the major roads. If a road is longer than 200 meters (current setting), we mark it a major road. If a road is connecting two major roads, we also mark it a major road.

**Figure 4**. The visualization of the simulated map. This is part of the Washington D.C. area.

## 2.2. Intelligent Driver Model

In Figure 4 we see the cars and taxis on the map. In order to simulate the traffic, we implemented the intelligent driver model (Intelligent driver model, n.d.) for all cars and taxis in the simulated environment. The driver will accelerate or decelerate a car according to the speed limit of a road that the car is moving on, the distance to the upcoming intersection, and the distance to the front car. When the car is far from its front car and the next intersection, it will be accelerated up to the speed limit using an accelerating factor equation. On the other hand, when the car is near its front car or the next intersection (if the car need to stop at the intersection), it will be decelerated by the same equation. The equation will calculate an accelerating factor and the speed difference is then calculated by the accelerating factor multiplied by the time period. The updated speed is then the original speed plus the newly calculated speed difference. The speed difference can be positive or negative representing acceleration or deceleration. If the traffic light at the next intersection shows the car can pass the intersection, then the distance to the intersection will not have an impact on the accelerating factor. The accelerating factor equation is shown in Figure 5. The equation for updating the speed and moving distance is shown in Figure 6.

```
Accelerating Factor Function:
    TIME_HEAD_AWAY = 1.5;        // unit: second
    DIST_GAP = 0.002;            // unit: kilometer (km)
    MAX_ACCELERATION = 0.001     // the maximum acceleration (km/s^2)
    MAX_DECELERATION = 0.003     // the maximum deceleration (km/s^2)

    // calculate the free road coefficient
    If front car exists: // the cars on the next lane can also be a front car
        frontCar = the front car
        distanceToFrontCar = the distance to the front car
        deltaSpeed = current car's speed – frontCar's speed
    Else:
        frontCar = None
        distanceToFrontCar = Infinity
        deltaSpeed = 0
    speedRatio = current car's speed / max speed of current car;
    freeRoadCoeff = pow(speedRatio, 4);

    // calculate the busy road coefficient
    timeGap = current car's speed * TIME_HEAD_AWAY / 3600;
    breakGap = current car's speed * deltaSpeed /
              (2 * math.sqrt(MAX_ACCELERATION * MAX_DECELERATION));
    safeDistance = DIST_GAP + timeGap + breakGap;
    If distanceToFrontCar > 0:
        distRatio = safeDistance / distanceToFrontCar;
        busyRoadCoeff = pow(distRatio, 2);
    Else:
        busyRoadCoeff = Infinity;

    // calculate the intersection coefficient
    safeIntersectionDist = 0.001 + timeGap + pow(current car's speed, 2) /
                          (2 * MAX_DECELERATION);
    distanceToStopLine = the distance to the stop line of the upcoming intersection;
    If distanceToStopLine > 0:
        safeInterDistRatio = safeIntersectionDist / distanceToStopLine;
        intersectionCoeff = pow(safeInterDistRatio, 2)
    Else:
        intersectionCoeff = Infinity

    // calculate the acceleration coefficient
    coeff = 1 - freeRoadCoeff - busyRoadCoeff - intersectionCoeff

    return round(MAX_ACCELERATION * coeff, 10)
```

**Figure 5.** The accelerating factor function that is used to control the speed of a car. The updated speed will be the original speed plus the multiplication of the accelerating factor with the time period.

```
Update Speed and Calculate Moving Distance:
    // get acceleration factor
    acceleration = get acceleration from "Accelerating Factor Function" in Figure 5

    // update current speed of the car
    currentSpeed += acceleration * second * 3600 // 1 hour has 3600 seconds
    currentSpeed = min(speedLimit, max(0, currentSpeed))

    // calculate the moving distance
    step = max(currentSpeed * second / 3600
            + 0.5 * acceleration * math.pow(second, 2), 0)
    If there is a front car:
        frontCarDistance = the distance to the front car
    Else:
        frontCarDistance = Infinity
    movingDistance = min(frontCarDistance, step) // cannot hit the front car
```

**Figure 6.** The equation for updating the speed and moving distance for a car.

In addition to the intelligent driver model that controls the speed of a car, the driver can also change to a quicker lane of the same road. If there is no other cars or taxis in front of the car that the driver is controlling, then the driver will not switch lanes. Otherwise, the driver will determine whether to switch the lanes according to the instance average speed of the lanes. This makes the simulation program more realistic.

## 2.3. Traffic Pattern for a Crash Event

To make the simulated environment more like a real world situation, when a car crashes, the program will reduce the speed limit of the road where a crash event happens to 10 km/hr. The crashed car will stop on the lane it moves on. However, other cars and taxis can still move on the other lane of the same road. The program will not manually interfere the traffic situation of other roads.

## 2.4. Navigation

For cars and taxis to arrive their destinations, we need a navigation system for them. The navigation system uses the average traffic time of roads to find the quickest route for every car and taxi. The calculation of the average traffic time will be described later. To find the quickest route, the system will perform the Dijkstra shortest path algorithm[7] and use the average traffic time (road length divided by the average speed) as the edge weight.

## 2.4.1. Average Traffic Time

To calculate the average traffic time, the system will keep tracking of the time that a car moves across a road. When a car enters a road (either from the start of the road or from a source point on a road), the system will create a record for the traffic time of the car moving on the road. The information included in the record are the starting position, current position, ending position, starting time, and ending time. In our setting, the system calculates the recent 5-minute average traffic time. The pseudo code for calculating the average road speed is shown in Figure 7.

---

[7] https://en.wikipedia.org/wiki/Dijkstra's_algorithm

**Average Traffic Time for A Road:**
/*
*Before calculate the average traffic time for a road, the system will first remove records whose ending time are more than a specific time (in our setting, the time is 5 minute) ago.*
*/
// *the shortest time for a car to drive across the entire road*
minTrafficTime = road.length / road.speedLimit * 3600

trafficTimeList = []
For record in records:  // *"records" holds all the remaining traffic time records*
   trafficTime = record.getTrafficTime(currentTime)
   **If** trafficTime is not None:
      trafficTimeList.append(trafficTime)

**If** trafficTimeList is not empty:
   averageTrafficTime = max(avg(trafficTime), minTrafficTime)
**Else:**
   // *No record means the road is empty in the past 5 minutes. So the car can drive at the*
   // *road's speed limit*
   averageTrafficTime = minTrafficTime

**return** averageTrafficTime

**record.getTrafficTime(currentTime):**
**If** record.endTime > 0:
   **If** record.endingPosition == record.startingPosition:
      **return** None
   **Else:**
      **return** (record.endingTime - record.startingTime) /
            (record.endingPosition - self.startingPosition)
**Else:**
   positionDiff = record.currentPosition - record.startingPosition
   **If** record.startingTime <= currentTime:
      timeDiff = currentTime - record.startingTime
   **Else:**
      timeDiff = globalTimeLimit - (record.startingTime - currentTime)

   **If** positionDiff == 0:
      **If** timeDiff == 0:  // *the car just entered this road, don't count this one*
         **return** None
      **Else:**
         /*
         *The car has entered this road but didn' make any move. So we approximate the*
         *traffic time by modifying the positionDiff. The modified positionDiff will be*
         *within the range 0.1~0.5. Then, the approximate traffic for this record is between*
         *2 to 10 times of the current traffic time that is timeDiff.*
         */
         positionDiff = min(0.5, max(0.1, 1 / timeDiff))

   **return** timeDiff / positionDiff

**Figure 7.** Pseudo code for calculating the average traffic time for roads.

## 3. Experiment

For this project, we experiment the use of navigation based on the average traffic time to find the quickest route for each car and taxi. We want to see whether the navigation system can help the called taxi to arrive first. Therefore, for comparison, two sets of experiments are performed: one of them will update the navigation for all the taxis every 30 seconds; another will only update the navigation for taxis when a new destination is assigned. In other words, we want to see if we keep updating the navigation periodically, can the called taxi arrive at the first order for more times than that the called taxi keep using the same navigation before it arrives its destination?

Total 30 experiments are performed and each of them user different seeds[8] for the random function and Poisson function in the program to fix the randomness for the same experiment. The setting and process for the experiments are listed below.

### 3.1. Simulation Setting
- The initial number of cars including the crashed car: 500.
- The initial number of taxis: 20.
- $\lambda$ for the Poisson arrival for general roads: 0.00002. [9]
- $\lambda$ for the Poisson arrival for major roads: 0.00006.
- The time between two consecutive loops in the simulator is 0.3 second.
- All random function is fixed by random seeds.
- Road speed limit: 60 km/hr.
- When a car crashed, the speed limit of the road will be set to 10 km/hr.
- Initially generate 80% cars on the major roads.
- The crash event will happen at the 5th minute after the simulation starts.
- The location of the crash event is fixed.
- The time interval for a traffic light controller to change its signals is between 15~30 seconds.
- For the experiment that the called taxi will update its navigation periodically, the time interval for an update is 30 seconds.

### 3.2. Simulation Process
- Every car and taxi will have a destination and it will find a routing based on the average traffic time when the destination is assigned.
- When a car arrives its destination, it will be removed from the map.
- When a taxi arrives its destination, a new destination will be assigned to it.
- New cars will be added to the map according to the Poisson point process at each source point.
- Before the crash event happens, when a taxi arrives its destination, a new destination will be assigned to that taxi.
- When a crash event happens, the simulator will call the "**nearest taxi**" (measured by average traffic time to the crash location) for the crash event and call rest of the taxis to go to the same location as well.
- When a taxi arrives the crash location, it will be removed from the map.

---

[8] The random seeds are: 2, 3, 5, 7, 11, 13, 17, 19, 23, 29, 31, 37, 41, 43, 47, 53, 61, 67, 71, 73, 79, 83, 89, 97, 101, 103, 107, 109, 113, 127.
[9] The program performs the Poisson point process at each source point every 0.3 second in our setting. So this $\lambda$ value should not be too large.

- Keep repeating the same steps above when the called taxi and at least 10 taxis have arrived the crash location.

### 3.3. Simulation Result

Figure 8 and Figure 9 show the distributions of the arrival order for the called taxi in 30 experiments under different navigation update policy.

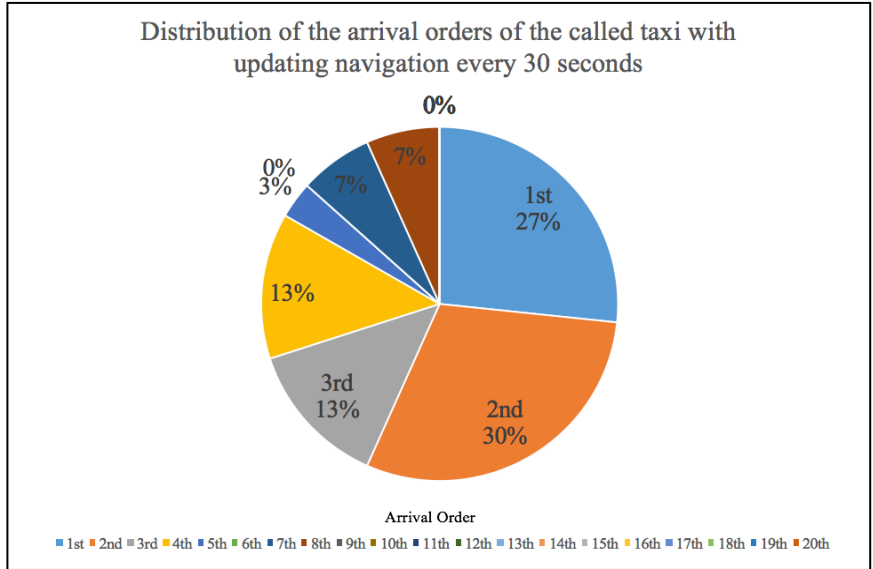

**Figure 8**. The distribution of the arrival order for the called taxi that updates its navigation every 30 seconds.

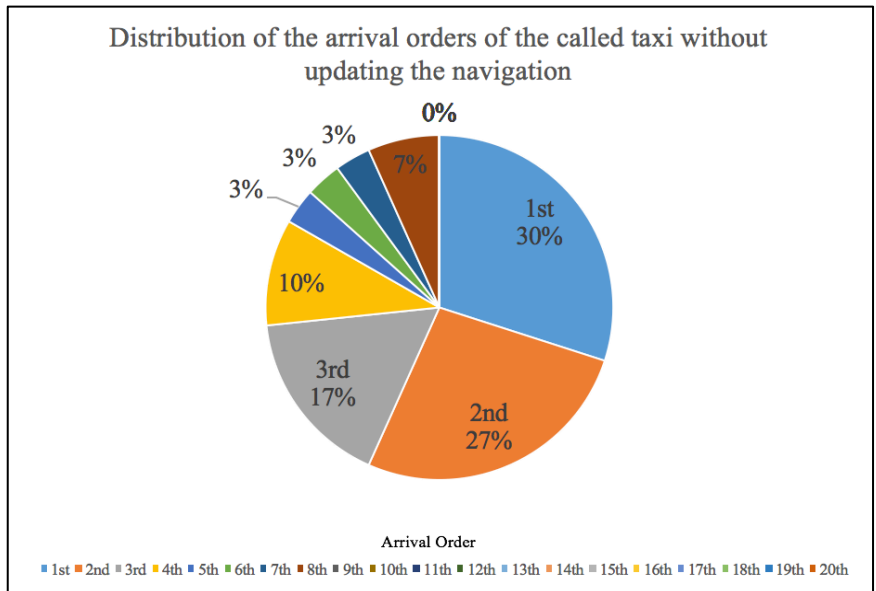

**Figure 9.** The distribution of the arrival order of the called taxi that doesn't update its navigation every 30 seconds.

# 4. Discussion

We want to test that if the called taxi updates its navigation periodically, it should arrive the crash location first. However, the experiment results show that the nearest taxi the system found when the crash event happens did not arrive first all the time. One reason is that using only the average traffic time is not enough to guarantee the first arrival for the nearest taxi. When the system finds the quickest route based on the average traffic time, the route might become slower later. On the other hand, the slower route might become quicker later. Although the nearest taxi didn't arrive the crash location first in all the experiments, it is still the first 3 taxis that arrive the crash location for at least 70% in all the experiments.

From the results, we know that using the navigation based on the average traffic time is not enough to guarantee the first arrival for the nearest taxi. The traffic pattern might change due to the crash event. So the average traffic time for the past 5 minutes cannot directly predict the traffic situation after the crash happened. It needs more advanced learning algorithms to help the system to predict the changes in the traffic pattern after a crash happens. The changes in the traffic pattern are hard to be learn using a supervised learning technique. We found an online learning algorithm called "MarcoPolo[10]" presented in (Arora, Dekel, & Tewari, Deterministic MDPs With Adversarial Rewards and Bandit Feedback, 2012) and (Arora, Dekel, & Tewari, Online Bandit Learning against an Adaptive Adversary: from Regret to Policy Regret, 2012) provided discussions about applying the online reinforcement learning algorithm for problems that have some similarities to our problem.

## 4.1. Possible Application of Online Reinforcement Learning

In the papers mentioned above, when the agent chooses an action at a state, then the adversary will give a reward to the agent according to the state-action information. The reward function is controlled by the adversary and will change over time (it may be affected by the actions that the agent chooses). One assumption is that the only feedback the agent observes is the reward it receives. This is also called "bandit feedback". Then the agent will adjust its policy at each round according to the reward it receives. It also assumes that the state transition is deterministic. That is, when every car chooses its action, the next state will be determined.

Back to the problem described in this report, we need two layers of this kind of online reinforcement learning algorithm: one is for the navigation to choose the quickest route and the other one is to choose the quickest taxi based on the policy learned from the environment when there is a crash event. The environment is the traffic situation and it will change over time. If we use an appropriate way to represent the state, the state space is large but finite. For example, we can use the number and position of taxis, the number of cars, the speed information, and other information of each road to represent a state. The action space is also finite. For the first algorithm, if the taxis choose an action at each intersection, the actions will be go straight, turn left, turn right, and etc. For the second algorithm, the action space will be the taxis. Which taxi should it call for the crash in one experiment?

We also need to design reward functions for the learning algorithms. Since we care the time for the called taxi to arrive the crash location, the reward function should be related to this factor. In our experiment, we also have the arrival order for all taxis. Therefore, the arrival order can also be a factor.

---

[10] **M**DP with **A**dversarial **R**ewards, weakly **CO**mmunicating structure, **P**artial feedback, by reduction to **O**nline **L**inear **O**ptimization.

If the MarcoPolo algorithm is applied to our problem appropriately with good definition of the state, action, and reward functions defined by ourselves, it should be able to improve the performance by predicting the changes in the traffic situation and make a better decision of which care should be called for a crash event.

## References

Arora, R., Dekel, O., & Tewari, A. (2012). *Deterministic MDPs With Adversarial Rewards and Bandit Feedback.* Proceedings of the Twenty-Eighth Conference on Uncertainty in Artificial Intelligence.

Arora, R., Dekel, O., & Tewari, A. (2012). *Online Bandit Learning against an Adaptive Adversary: from Regret to Policy Regret.* Proceedings of the 29th International Conference on Machine Learning.

Bhalla, K. (2015). *Taxi-based Ambulance.* Personal Communication.

*Intelligent driver model*. (n.d.). Retrieved from Wikipedia: https://en.wikipedia.org/wiki/Intelligent_driver_model

Sutton, R., & Barto, A. (1998). *Reinforcement Learning: An Introduction.* Cambridge, MA: MIT Press.

*TRANSIMS*. (n.d.). Retrieved from U.S. Department of Transportation Federal Highway Administration: http://www.fhwa.dot.gov/planning/tmip/resources/transims/

*Transims Open Source*. (n.d.). Retrieved from Google Code: https://code.google.com/archive/p/transims/

Volkhin, A. (n.d.). *Road Traffic Simulator*. Retrieved from Github: https://github.com/volkhin/RoadTrafficSimulator